\documentclass[conference,harvard,brazil,english]{sbatex}
\usepackage[latin1]{inputenc}
\usepackage{ae}

\makeatletter
\def\verbatim@font{\normalfont\ttfamily\footnotesize}
\makeatother
\usepackage{amsmath}

\makeatother
\usepackage{amsmath}
\usepackage{longtable}
\usepackage{bm}
\usepackage{afterpage} 
\usepackage{float} 
\usepackage[normalem]{ulem}
\usepackage{amsfonts} 
\usepackage{graphicx,adjustbox} 
\usepackage{color}

\begin{document}


\title{Computation of Extended Robust Kalman Filter for Real-Time Attitude and Position Estimation}

\author{Gaurav R. Yengera}{g.yengera@gmail.com}
\address{ Department of Electrical Engineering, Indian Institute of Technology (Banaras Hindu University), Varanasi and also with Department of Electrical Engineering, University of S\~ao Paulo at S\~ao Carlos, S\~ao Paulo, Brazil}

\author{Roberto S. Inoue}{rsinoue@ufscar.br}
\address{Department of Electrical Engineering, Federal University of S\~ao Carlos at S\~ao Carlos, S\~ao Paulo, Brazil}

\author{Mundla Narasimhappa}{mr.narasimha08@gmail.com}
\address{Department of Electrical Engineering, University of S\~ao Paulo at S\~ao Carlos, S\~ao Paulo, Brazil}

\author[3]{Marco H. Terra}{terra@sc.usp.br}


\twocolumn[

\maketitle

\selectlanguage{english}
\begin{abstract}
 This paper deals with the implementation of the extended robust Kalman filter (ERKF) which was developed considering uncertainties in the parameter matrices of the underlying state-space model. A key contribution of this work is the demonstration of a method for real-time computation of the filter on parallel computing devices. The solution of the filter is expressed as a set of simultaneous linear equations, which can then be evaluated based on QR decomposition using Givens rotation. This paper also presents the application of the ERKF in the development of an attitude and position reference system for a cargo transport vehicle. This work concludes by analyzing the performance of the ERKF and verifying the validity of the Givens rotation method.
\end{abstract}



]


\selectlanguage{english}
\section{Introduction}

The Kalman filter \cite{kalman60a,Anderson1979,kailath_2000} has played an important role in solving estimation problems appearing in navigation, economics, communications, control and other areas. The real-time computation of the Kalman filter has been an important and recurring feature in many of its applications. A fundamental assumption in the Kalman filter is that the underlying state-space model is accurate and does not contain uncertainties. When this condition is violated, the performance of the filter could deteriorate drastically \cite{sayed2001}. As a result robust estimation is necessary in several real world applications and certain algorithms have been developed for this purpose. However, very little research has been carried out on the real time implementation of these robust estimation algorithms on parallel computing devices such as FPGAs and GPUs.

A detailed explanation of the robust optimal filtering approach utilized in the ERKF has been presented in \cite{Ishihara2015,Inoue2016}. Also its advantages over other robust filtering approaches have been discussed in those papers. To summarize the important features of the ERKF: it does not require any auxiliary parameter to be tuned while for linear systems, stability and convergence are guaranteed for all steady-state estimates. Thereby this filtering approach is preferable for real time implementation as no offline computations are necessary. Additionally the ERKF assumes the existence of uncertainties in all parameter matrices of the state-space model. 

FPGAs have been a popular choice for the real time implementation of the Kalman filter. The most significant prior work on floating-point FPGA implementations have been developed based on direct mapping of the equations on the FPGA which either involves explicit matrix inversion, see for instance \cite{Bonato2007,Lee1997}, or representation of the equations in Schur complement form and then applying Fadeev's algorithm as done in  \cite{Chen2005}. 

This paper will focus on the implementation of the ERKF presented in \cite{Inoue2016}. For the evaluation of the ERKF, a matrix of particularly large dimensions needs to be inverted. Hence, the matrix inversion approach would be computationally intensive and not ideal for real time applications. It will be shown that the proposed method is computationally more efficient than the conventional matrix inversion approach. As compared to Fadeev's algorithm, it is more  straightforward to evaluate the filter as a set of simultaneous linear equations by applying QR decomposition using Givens rotation, see  \cite{Francis1962,Givens1958}, followed by back substitution to obtain desired state vector and state covariance matrix.

The application of the ERKF in an attitude and position reference system for a cargo transport vehicle  utilizing a global positioning system (GPS) along with an inertial measurement unit (IMU) is presented. To accommodate for the lower measurement update frequency of the GPS, as discussed in \cite{farrel2008}, an inertial navigation system based on attitude estimates is used to estimate position in the absence of GPS measurements. At the same time the IMU measurements contain residual errors even after calibration and the ERKF is used to ensure that state estimates are robust to such errors or uncertainties. In \cite{Inoue2016} the performance improvement of the ERKF over the extended Kalman filter in the presence of uncertainties has been discussed. The role of the ERKF is especially crucial when working with low-cost IMUs which tend to possess substantial uncertainties 
. A method to model the uncertainties present in the IMU rate gyros and accelerometers measurements, and incorporate them into the ERKF has been shown. This application is of general importance for developing more sophisticated navigation systems and it also highlights the necessity for real time computation of the ERKF. This paper concludes by using sensor data collected from this experimental setup to verify the Givens rotation based computation approach.  

The organization of the paper is as follows: the extended robust Kalman filter is presented in Section \ref{sec:ERKF}. In Section \ref{sec:CGR}, the proposed algorithm and its computational complexity are discussed. The vehicle attitude and position estimation system is presented in Section \ref{sec:systems}. In Section \ref{sec:results}, the experimental setup and results are discussed. Finally, Section \ref{sec:conclu} presents the conclusion of the paper.

\section{Extended Robust Kalman Filter} \label{sec:ERKF}

For the development of the robust Kalman filter, the underlying state-space model of the dynamic discrete-time system is modified to incorporate parametric uncertainties:
\small
\begin{align}\label{state_space}
\begin{split}
\textbf{x}_{k+1}&=\left(F_k+\delta F_k\right)\textbf{x}_k+ \left(G_{k} + \delta G_{k}
\right)\textbf{w}_{k},
\\
\textbf{z}_{k}&=\left(H_{k}\!+\!\delta H_{k}\right)\textbf{x}_{k}\! +\!
\left( K_{k} \!+\! \delta K_{k} \right) \textbf{v}_{k},
\end{split}
\end{align}
\normalsize for $k\geq0$, where $\textbf{x}_k \in \mathbb{R}^n$ is the state vector,
$\textbf{z}_{k} \in \mathbb{R}^p$ is the measurement vector, $\textbf{w}_k$ and $\textbf{v}_k$ are the noise vectors corresponding to the state update and measurement equations respectively. Typically, $\textbf{x}_{0}$, $\textbf{w}_{k}$ and $\textbf{v}_{k}$ are considered as mutually independent zero-mean Gaussian random variables with respective variances $\mathbb{E}\{\textbf{x}_{0}\textbf{x}_{0}^{T}\} =\varPi_{0} \ge 0$, $\mathbb{E}\{\textbf{w}_{k}\textbf{w}_{k}^{T}\} = Q_{k} \ge 0$ and $\mathbb{E}\{\textbf{v}_{k}\textbf{v}_{k}^{T}\} = R_{k} \ge 0$.  $F_k\in\mathbb{R}^{n\times n}$, $G_k\in\mathbb{R}^{n\times{m}}$,
$H_{k}\in\mathbb{R}^{p\times{n}}$,
$K_{k}\in\mathbb{R}^{p\times{m}}$ are nominal parameter
matrices,  and  $\delta F_k\in\mathbb{R}^{n\times n}$, $\delta
G_k\in\mathbb{R}^{n\times{m}}$, $\delta
H_{k}\in\mathbb{R}^{p\times{n}}$, $\delta
K_{k}\in\mathbb{R}^{p\times{m}}$ are uncertainty matrices which are modeled as:
\small
\begin{align}\label{incertezas}
\begin{bmatrix}
\delta F_k & \delta G_k
\end{bmatrix}&=&M_{1_{k}}\Delta_1 \begin{bmatrix} N_{F_{k}} & N_{G_{k}} \end{bmatrix} \\
\begin{bmatrix}
\delta H_k & \delta K_k
\end{bmatrix}&=&M_{2_{k}}\Delta_2 \begin{bmatrix} N_{H_{k}} & N_{K_{k}} \end{bmatrix}
\end{align}

\normalsize where $||\Delta_1||<1$ and $||\Delta_2||<1$ are arbitrary contractions. Matrices $M_{1_{k}}$, $M_{2_{k}}$, $N_{F_{k}}$, $N_{G_{k}}$, $N_{H_{k}}$, and $N_{K_{k}}$ are known \textit{a-priori}. 

The ERKF is developed considering the solution of the following unconstrained optimization problem:
\begin{eqnarray}\label{eou}&&
\min_{\textbf{x}_k,\textbf{x}_{k+1}}\max_{\delta_k} \{\mathcal{J}^\mu_k(\textbf{x}_k,\textbf{w}_k,\textbf{v}_{k},\textbf{x}_{k+1},\delta_k)\},
\end{eqnarray}
\normalsize  where $\delta_k:=\{\delta F_k,\delta G_k,\delta
K_{k},\,\delta H_{k}\}$. The cost function $\mathcal{J}^\mu_k(\textbf{x}_k,\textbf{w}_k,\textbf{v}_{k},\textbf{x}_{k+1},\delta_k)$ is given in \cite{Inoue2016}.

%
%

\begin{table*}
	\caption{Recursive Algorithm for Extended Robust Kalman filter}
	\label{table} \centering
	\begin{tabular}{|l|}           
		\hline
		\scriptsize \textbf{Uncertain Model:} Consider (\ref{state_space}) with $\varPi_{0}\succ{0}$, $Q_{k}\succ{0}$, and $R_{k}\succ{0}$. \\
		\scriptsize \textbf{Step $0$:} (Initial Conditions) $P_{0|-1}\,=\,\varPi_{0}$, $\widehat{\textbf{x}}_{0|-1}\,=\,0$. \\
		\scriptsize \textbf{Step $k$:} Given $\textbf{z}_{k}$, update $\{\widehat{\textbf{x}}_{k|k}\,;\,\widehat{\textbf{x}}_{k+1|k}\,;\,P_{k+1|k}\}$ from $\{\textbf{z}_{k}\,;\,\widehat{\textbf{x}}_{k|k-1}\,;\,P_{k|k-1}\}$ as follows: \\
		\begin{scriptsize}
			$\hfill\begin{bmatrix}
			\widehat{\textbf{x}}_{k|k}   & \ast \\
			\widehat{\pmb{\nu}}_{k|k}    & \ast \\
			\widehat{\textbf{x}}_{k+1|k} & P_{k+1|k} \end{bmatrix} =
			\begin{bmatrix}
			\widehat{\textbf{x}}_{k|k-1} & 0 \\
			0 & 0 \\
			0 & 0 \end{bmatrix} + \setcounter{MaxMatrixCols}{7}\begin{bmatrix}
			0&0&0&0&I&0&0\\0&0&0&0&0&I&0\\0&0&0&0&0&0&I\end{bmatrix}\hfill$
		\end{scriptsize}\\
		\begin{scriptsize}
			$\hfill\setcounter{MaxMatrixCols}{7}\begin{bmatrix}
			P_{k|k-1}&0&0&0&I&0&0\\
			0&\mathcal{R}_{k}&0&0&0&I&0\\
			0&0&0&0&\mathcal{F}_{k}&\mathcal{G}_{k}&\mathcal{E}_{k}\\
			0&0&0&0&N_{\mathcal{F}_k}&N_{\mathcal{G}_k}&N_{\mathcal{E}_k}\\
			I&0&\mathcal{F}^T_{k}&N^T_{\mathcal{F}_k}&0&0&0\\
			0&I&\mathcal{G}^T_{k}&N^T_{\mathcal{G}_k}&0&0&0\\
			0&0&\mathcal{E}^T_{k}&N^T_{\mathcal{E}_k}&0&0&0
			\end{bmatrix}\setcounter{MaxMatrixCols}{10}
			^{-1}
			\begin{bmatrix}
			0 & 0 \\
			0 & 0 \\
			\textbf{b}_{k} & 0 \\
			N_{\textbf{b}_{k}} & 0 \\
			0 & 0 \\
			0 & 0 \\
			0 & -I
			\end{bmatrix},\hfill$
		\end{scriptsize}
		
		\\
		
		%
		
		\begin{scriptsize}
			$\hfill \widehat{\pmb{\nu}}_{k|k} = \begin{bmatrix} \widehat{\textbf{w}}_{k|k} \\ \widehat{\textbf{v}}_{k|k} \end{bmatrix}$, $\mathcal{F}_{k} = \begin{bmatrix} F_{k} \\ H_{k} \end{bmatrix}$, $\mathcal{G}_{k} = \begin{bmatrix} G_{k} & 0 \\ 0 & K_{k} \end{bmatrix}$, $\mathcal{E}_{k} = \begin{bmatrix} -I \\ 0 \end{bmatrix},\hfill$
		\end{scriptsize}
		
		\\
		
		\begin{scriptsize}
			$\hfill N_{\mathcal{F}_{k}} = \begin{bmatrix} N_{F_{k}} \\ N_{H_{k}} \end{bmatrix}$, $N_{\mathcal{G}_{k}} = \begin{bmatrix} N_{G_{k}} & 0 \\ 0 & N_{K_{k}} \end{bmatrix}$, $N_{\mathcal{E}_{k}} = \begin{bmatrix} 0 \\ 0 \end{bmatrix},\hfill$
		\end{scriptsize}
		
		\\
		
		\begin{scriptsize}
			$\hfill \textbf{b}_{k} = \left [ \begin{array}{c} -F_{k} \widehat{\textbf{x}}_{k|k-1} \\ \textbf{z}_{k} - H_{k}\widehat{\textbf{x}}_{k|k-1}\end{array} \right]$, $N_{\textbf{b}_{k}} = \left [ \begin{array}{c} - N_{F_{k}} \widehat{\textbf{x}}_{k|k-1} \\ - N_{H_{k}} \widehat{\textbf{x}}_{k|k-1}\end{array} \right]$, $\mathcal{R}_{k} = \begin{bmatrix} Q_{k} & 0 \\ 0 & R_{k} \end{bmatrix}$
		\end{scriptsize}
		
		\\
		
		\hline
	\end{tabular}
\end{table*} 

The ERKF is essentially in the form of a predicted estimator filter where the measurement update is computed first and the time update step later. Thereby the filter recursively computes $\widehat{\textbf{x}}_{k+1|k}$ and $P_{k+1|k}$ from $\widehat{\textbf{x}}_{k|k-1}$ and $P_{k|k-1}$ respectively.

The optimal estimates for linear systems are obtained by substituting $\mu \rightarrow \infty$. The recursive algorithm for obtaining the optimal filtered estimates is presented in Table \ref{table}. 

The ERKF is applied to nonlinear system models by defining $\textbf{b}_{k} = \begin{bmatrix}
-F_{k}\widehat{\textbf{x}}_{k|k-1} \\ \textbf{z}_{k} - \textbf{h}(\widehat{\textbf{x}}_{k|k-1})
\end{bmatrix}$ and is used with linear system models by defining $\textbf{b}_{k} = \left [ \begin{array}{c} -F_{k} \widehat{\textbf{x}}_{k|k-1} \\ \textbf{z}_{k} - H_{k}\widehat{\textbf{x}}_{k|k-1}\end{array} \right]$.

\section{Computation using Givens Rotation} \label{sec:CGR}


The ERKF given in Table \ref{table} can be rewritten as a linear system $A\textbf{y}=\textbf{b}$, as shown in (\ref{sim_eqns}). The $\mu$ and $\lambda$ terms are values we are not interested in, while all the other terms are as defined in Table \ref{table}. 

\begin{equation}
\scriptsize
\resizebox{0.47\textwidth}{!}{$
	\hfill\setcounter{MaxMatrixCols}{7}\underbrace{\begin{bmatrix}
		P_{k|k-1}&0&0&0&I&0&0\\
		0&\mathcal{R}_{k}&0&0&0&I&0\\
		0&0&0&0&\mathcal{F}_{k}&\mathcal{G}_{k}&\mathcal{E}_{k}\\
		0&0&0&0&N_{\mathcal{F}_k}&N_{\mathcal{G}_k}&N_{\mathcal{E}_k}\\
		I&0&\mathcal{F}^T_{k}&N^T_{\mathcal{F}_k}&0&0&0\\
		0&I&\mathcal{G}^T_{k}&N^T_{\mathcal{G}_k}&0&0&0\\
		0&0&\mathcal{E}^T_{k}&N^T_{\mathcal{E}_k}&0&0&0\\
		\end{bmatrix}}_{A_{1}}\setcounter{MaxMatrixCols}{7}\underbrace{
		\begin{bmatrix}
		\lambda_{1}&\mu_{1}\\
		\lambda_{2}&\mu_{2}\\
		\lambda_{3}&\mu_{3}\\
		\lambda_{4}&\mu_{4}\\
		\widehat{\textbf{x}}_{k|k} - \widehat{\textbf{x}}_{k|k-1}  & \ast \\
		\widehat{\pmb{\nu}}_{k|k}    & \ast \\
		\widehat{\textbf{x}}_{k+1|k} & P_{k+1|k} \end{bmatrix}}_{\textbf{y}_{1}} = 
	\underbrace{\begin{bmatrix}
		0 & 0 \\
		0 & 0 \\
		\textbf{b}_{k} & 0 \\
		N_{\textbf{b}_{k}} & 0 \\
		0 & 0 \\
		0 & 0 \\
		0 & -I
		\end{bmatrix}}_{\textbf{b}_{1}},\label{sim_eqns}\hfill$}
\end{equation}

To solve for only the required elements of matrix $\textbf{y}$, i.e. state vector $\widehat{\textbf{x}}_{k+1|k}$ and state covariance matrix $P_{k+1|k}$, each system of linear equations corresponding to an individual column of matrix $\textbf{y}$ is evaluated one at a time using QR decomposition based on Givens rotation. And only the required values are calculated using back-substitution. The following equations describe the proposed solution:
\begin{equation}
A\textbf{y}_{l} = \textbf{b}_{l},
\end{equation}
\begin{equation}
QR\times \textbf{y}_{l} = b_{l},
\end{equation}
\begin{equation}
R\times \textbf{y}_{l} = Q^{-1}\times b_{l} = Z_{l},\label{backsusbstitution2}
\end{equation}
where $\textbf{y}_{l}$ and $b_{l}$ are the $l$th columns of $\textbf{y}$ and $\textbf{b}$ respectively, and $A = QR$ is the result of QR decomposition of matrix $A$. 

The matrices $R$ and $Z_{l}$ can be obtained by the following equation:
\begin{equation}
[R,Z_{l}] = \Theta[A, b_{l}] = \Theta M_{l},\label{givens2}
\end{equation}
where $\Theta$ is any sequence of Givens rotations which result in the upper triangularization of matrix $A$. Since matrix $R$ is a triangular matrix, the elements of vector $\textbf{y}_{l}$, specifically the ones corresponding to state vector $\widehat{\textbf{x}}_{k+1|k}$ or state covariance matrix $P_{k+1|k}$, can be calculated from (\ref{backsusbstitution2}) using back-substitution. All the elements of $\widehat{\textbf{x}}_{k+1|k}$ and $P_{k+1|k}$ are obtained after (\ref{givens2}) and (\ref{backsusbstitution2}) have been carried out for all the columns of $\textbf{y}$.

The number of floating point operations (FLOPS) required by this method, where one FLOP is counted as any individual floating point operation; has been calculated referring to \cite{matrixcomp} and considering the following general case: $\textbf{x}\in \mathbb{R}^{n\times 1}$, $P\in\mathbb{R}^{n\times n}$, $A\in\mathbb{R}^{m\times m}$. It can be noticed that in (\ref{sim_eqns}), the matrix $A$ is a square matrix and $m \gg n$. Thereby, $\textbf{y}\in\mathbb{R}^{m\times (n+1)}$, $\textbf{b}\in\mathbb{R}^{m\times (n+1)}$ and $M_{l}\in\mathbb{R}^{m\times (m+1)}$. 

Givens rotation, in (\ref{givens2}), requires FLOPS of an order of magnitude equal to $3(m+1)^{2}(m - \frac{m+1}{3})$. Since $m \gg 1$, this can be approximated as $2m^{3}$. Back-substitution is applied in (\ref{backsusbstitution2}) to calculate the bottom $n$ elements of vector $\textbf{y}_{l}$ as only these correspond to elements of either $\widehat{\textbf{x}}_{k+1|k}$ or $P_{k+1|k}$. Hence, the FLOPS required in (\ref{backsusbstitution2}) is of the order of magnitude $n^{2}$.

Remembering that (\ref{givens2}) and (\ref{backsusbstitution2}) need to be carried out $n+1$ times, once for each column of matrix $\textbf{y}$, the total number of FLOPS required is given by:
\begin{equation}
\text{Total FLOPS} = (n+1)\times (2m^{3} + n^{2}) \propto 2nm^{3}.\label{flops_givens}
\end{equation}

For evaluating the ERKF directly as presented in Table \ref{table}, explicitly calculating the inverse of matrix $A$ using Gaussian elimination would involve applying Gaussian elimination and back-substitution $m$ times. The number of FLOPS required would be of the order of magnitude $\frac{2m^{4}}{3}$. To only obtain $\widehat{\textbf{x}}_{k+1|k}$ and $P_{k+1|k}$ after having obtained the inverse, requires matrix multiplication considering the bottom $n$ rows of $A^{-1}$ alone. The number of FLOPS for the matrix multiplication step would be $2nm(n+1)$. The computationally intensive step in this method is the matrix inversion step.

Comparing the computational cost of the matrix inversion approach, $\frac{2m^{4}}{3}$, with (\ref{flops_givens}) and noticing from the structure of matrix $A$ that $n < \frac{m}{3}$; it can be concluded that the Givens rotation approach is computationally more efficient. 

An important remark to be added here is that instead of the QR decomposition approach, the more efficient LU decomposition approach to solve the linear system given in (\ref{sim_eqns}) would appear to be a better choice. However, it was seen that the result of the LU decomposition method was unstable and did not converge with the results obtained from the matrix inversion approach. It can further be noticed that matrix $A$ is sparse and that is the reason for Givens rotation being preferred over Householder reflections. Additionally, it is important to note that Givens rotation based QR decomposition easily lends itself to parallel implementations \cite{Wang:2009}. Hence, the computation speed can further be increased on a parallel computing device such as an FPGA. 

\section{Vehicle State Determination}\label{sec:systems}

\begin{figure}[ht]
	\centering
	\includegraphics[width=0.8\linewidth]{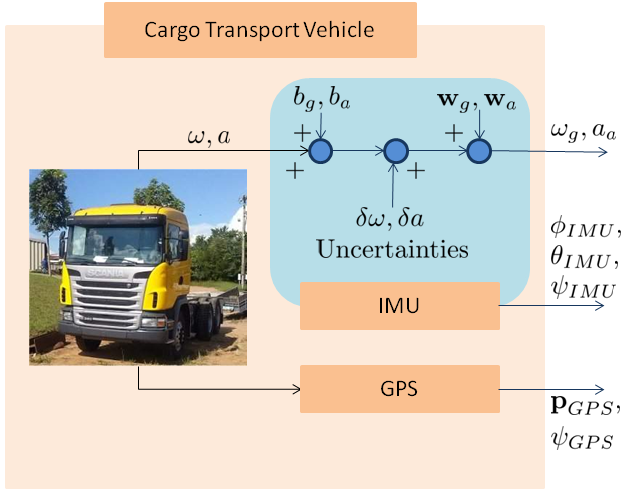} 
	\caption{Cargo transport vehicle experimental setup.} \label{fig:diagram_scania}
\end{figure}

In this section the vehicle state determination system (attitude, position and velocity) is presented. The system is composed of an IMU and a GPS module. The IMU is based on uncertain output models of rate gyros and accelerometers $\bm{\omega}_{\text{g}}$, $\bm{a}_{\text{a}}$. In Figure \ref{fig:diagram_scania} the model of the IMU considers rate gyro bias $\bm{b}_{\text{g}}$, accelerometer bias $\bm{b}_{\text{a}}$, Gaussian white noise in the rate gyros and accelerometers, $\textbf{w}_{\text{g}}$ and $\textbf{w}_{\text{a}}$, respectively, and uncertain terms due to scale factor and axes misalignment of the rate gyros, accelerometers, $\delta\bm{\omega}_{\text{g}}$ and $\delta\bm{a}_{\text{a}}$, respectively. The IMU also measures the tilt angles $\phi_{IMU}$ and $\theta_{IMU}$, as well as the yaw angle $\psi_{IMU}$. The GPS module provides geodetic position $\bm{p}_{GPS}$ and yaw angle $\psi_{GPS}$.

\begin{figure}[ht]
	\centering
	\includegraphics[width=0.8\linewidth]{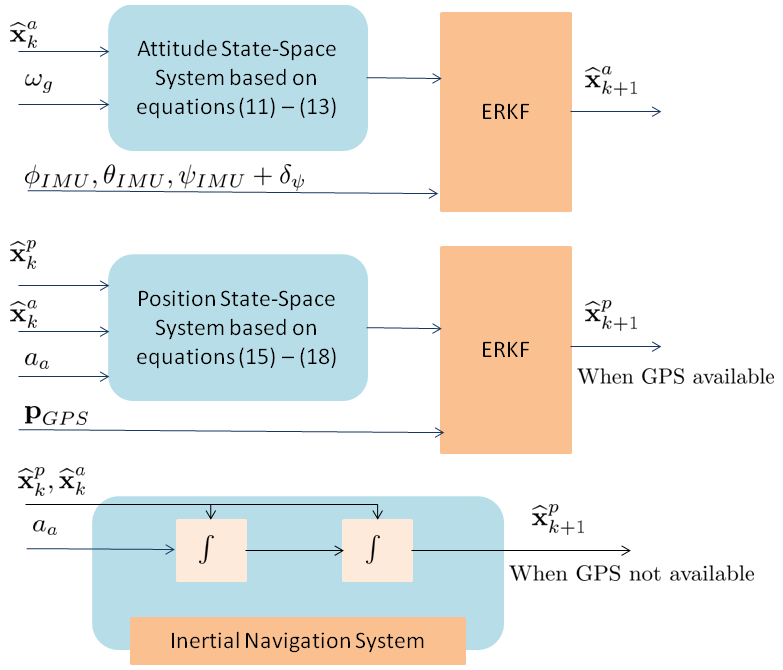} 
	\caption{Estimation Diagram.} \label{fig:diagram_systems}
\end{figure}

The estimation of the state is split in two filters: (1) an attitude estimator and (2) a position estimator, in order to obtain a trade-off between accuracy and processing power \cite{bijker2008}. In Figure \ref{fig:diagram_systems}, the method of obtaining attitude and position estimates has been illustrated. This is essentially a pictorial representation of the systems described in Sections \ref{sec:attitude} and \ref{sec:position}. Further, Figure \ref{fig:diagram_systems} shows how the inertial navigation system updates position estimates in the absence of GPS measurements by integrating accelerometer readings. The states of the attitude and position system are represented by $\widehat{\textbf{x}}^{a}$, and $\widehat{\textbf{x}}^{p}$, respectively.


\subsection{Attitude system}\label{sec:attitude}

The dynamic equations of the attitude model are given by \cite{farrel2008,kfbeginner}:
\begin{align}
[\dot{\phi}~~\dot{\theta}~~\dot{\psi}]^{T}&=\Omega(\phi, \theta, \psi)[p~~q~~r]^T, \label{eq:angularvelocitiesrelation} \\
[p~~q~~r]^T  &=  \bm{\omega}_{\text{g}}  + \delta\bm{\omega}_{\text{g}}-\bm{b}_{\text{g}} - \textbf{w}_{\text{g}},\label{eq:rategyromodel} \\
\dot{\bm{b}}_{\text{g}} &= -\frac{1}{\tau_{\text{g}}}\bm{b}_{\text{g}} + \textbf{w}_{b_{\text{g}}}, \label{eq:rategyrobias}
\end{align}
where $\phi$ and $\theta$ are the tilt angles roll and pitch, respectively; $\psi$ is the yaw angle, $p$, $q$ and $r$ are the angular velocities in the body frame; $\tau_{\text{g}}$ is the correlation time of the Gauss Markov process; $\textbf{w}_{b_{\text{g}}}$ is Gaussian white noise of the rate gyros bias; and 
$\Omega(\phi, \theta, \psi)$ is the transformation matrix between angular velocities,  which is given by:
\begin{align}
\Omega(\phi, \theta, \psi) = \left[ \begin{array}{ccc}
1 & \sin\phi \tan\theta & \cos\phi\tan\theta \\
0 & \cos\phi 				& -\sin\phi\\
0 & \sin\phi\sec\theta  & \cos\phi\sec\theta\end{array} \right].
\end{align}

The filters presented in this paper are based on discrete-time systems. In this regard, Equations (\ref{eq:angularvelocitiesrelation}) - (\ref{eq:rategyrobias}) are represented in the form of (\ref{state_space}) after being linearized and discretized considering a sample time $T$. The terms of (\ref{state_space}) are chosen as; $\textbf{x}^{a} = [\phi~~\theta~~\psi~~\bm{b}_{\text{g}}^T]^{T}\in \mathbb{R}^{6 \times 1}$ is the state vector,  $\textbf{w}^{a} = [\textbf{w}_{\text{g}}^T ~\textbf{w}_{b_{\text{g}}}^T]^{T}\in \mathbb{R}^{6 \times 1}$ is the vector Gaussian process with zero mean and covariance $Q^a$, $\textbf{z}^{a} = [\phi_{IMU}~~\theta_{IMU}~~\psi_{IMU}+\delta_{\psi}]^{T} \in \mathbb{R}^{3 \times 1}$ is the measurement vector, $\delta_{\psi} = \psi_{GPS} - \psi_{IMU}$ is the yaw error between GPS and IMU computed when GPS is available, $\textbf{v}^{a} \in \mathbb{R}^{3\times 1}$ is the vector Gaussian process with zero mean and covariance $R^a$ of the measured angles in $\textbf{z}^{a}$, $F_{k}^a$ is the state transition matrix, $G_{k}^a$ is the input noise matrix, and $H_k ^a= [I_{3 \times 3} ~~0_{3 \times 3}]$ is the measurement matrix. 


\subsection{Position system}\label{sec:position}

The dynamic equations of the position model are given by \cite{farrel2008,bijker2008}:
\begin{align}
[ \dot{\lambda} ~~\dot{\varphi} ~~ \dot{h}] ^{T} &= \Psi(\lambda,\varphi,h)
[ \upsilon_{N} ~~ \upsilon_{E} ~~ \upsilon_{D} ]^{T},			\label{eq:vargeod}  \\
[ \dot{\upsilon}_{N} ~~ \dot{\upsilon}_{E} ~~ \dot{\upsilon}_{D} ] ^{T}
&= \bm{g}_{\text{e}} + A^{T}(\phi, \theta, \psi) \bm{a} 
, \label{eq:velNED01}\\						
\bm{a} &=  \bm{a}_{\text{a}} + \delta \bm{a}_{\text{a}}- \bm{b}_{\text{a}}  - \textbf{w}_{\text{a}}, \label{eq:accelmodel}\\
\dot{\bm{b}}_{\text{a}} &= -\frac{1}{\tau_{\text{a}}}\bm{b}_{\text{a}} + \textbf{w}_{b_{\text{a}}}, \label{eq:biasacel}
\end{align}
where $\bm{p} = [\lambda \;\; \varphi \;\; h]^{T}$ are geodetic positions in the LLA (Latitude, Longitude and Altitude) frame ; 
$\bm{\upsilon} = [\upsilon_{N} \;\; \upsilon_{E} \;\; \upsilon_{D}]^{T}$ are the velocities in the NED (\textit{North}, \textit{East} and \textit{Down}) frame;
$R_{\lambda}$ is the radius of meridian curvature at a given latitude;
$ R_{\phi}$ is the transverse radius of curvature,
$\bm{g}_{\text{e}}$ is the Earth's gravity vector; 
$\bm{a}$ is the actual linear acceleration;
$A(\phi, \theta, \psi)$ is the rotation matrix from Inertia frame to Body frame;
$\tau_{\text{a}}$ is the correlation time of the Gauss Markov process; $\textbf{w}_{b_{\text{a}}}$ is Gaussian white noise of the accelerometer bias; and 
$\Psi(\lambda,\varphi,h)$ is the transformation matrix between linear velocities,  which is given by:
\begin{align}
\Psi(\lambda,\varphi,h) =
\left [ \begin{array}{ccc} \frac{1}{R_{\lambda} + h} & 0 																		& 0 \\
0												 & \frac{1}{(R_{\phi} + h) cos \lambda} & 0 \\
0												 & 0 																		& -1 \end{array} \right].
\end{align}

Equations (\ref{eq:vargeod})-(\ref{eq:biasacel}) are written in the state-space form of (\ref{state_space}) after being linearized and discretized, such that state vector $\textbf{x}^{p} = [\bm{p}^T~~\bm{\upsilon}^T~~\bm{b}_{\text{a}}^T]^{T}\in \mathbb{R}^{6 \times 1}$,  $\textbf{w}^{p} = [\textbf{w}_{\text{a}}^T ~\textbf{w}_{b_{\text{a}}}^T]^{T}\in \mathbb{R}^{6 \times 1}$ is the vector Gaussian process with zero mean and covariance $Q^p$, $\textbf{z}^{p} = [\bm{p}_{GPS}^T]^{T} \in \mathbb{R}^{3 \times 1}$ is the measurement vector, $\textbf{v}^{p} \in \mathbb{R}^{3\times 1}$ is the vector Gaussian process with zero mean and covariance $R^p$ of the measured position and velocity in $\textbf{z}^{p}$, $F_{k}^p$ is the state transition matrix, $G_{k}^p$ is the input noise matrix, and $H_k^p = [I_{3 \times 3} ~~0_{3 \times 6}]$ is the measurement matrix. 


\section{Experimental Results}\label{sec:results}

\subsection{Description of Experimental Setup}

An IMU and GPS were used to track the attitude and position of the cargo transport vehicle shown in Figure \ref{fig:diagram_scania}. The IMU used was the Xsens MTi-300-AHRS-2A5G4, which utilized MEMS based sensors. It comprised of a 3-axial accelerometer, 3-axial gyroscope and a 3-axial magnetometer. The update rate of the IMU was 400 Hz and it had an in-built algorithm which computed orientation, angular velocity and linear velocity from sensor readings. For this experimental setup, orientation was measured in terms of Euler angles and both the angular and linear velocities were measured about the body reference frame of the vehicle.


Septentrio AsteRx2eH PRO GPS was used in the cargo transport vehicle. The update rate of the GPS was 10 Hz and this was considerably slower than the IMU update rate. The measurements provided by the GPS were latitude, longitude, altitude, linear velocity as well as heading or yaw angle.

The weighting matrices $Q^s$ and $R^s$ for the ERKF were chosen based on the method described in \cite{xing2008}, where $s = a,~p$.  And the parameter matrices $N_{F_{k}}^s$, $N_{G_{k}} ^s$,  and $N_{H_{k}}^s$ are modeled in a manner to attenuate the uncertain terms $\delta\bm{\omega}_{\text{g}}$ presented in (\ref{eq:rategyromodel}), $\delta \bm{a}_{\text{a}}$ presented in (\ref{eq:accelmodel})  and others sources of uncertainties occurring in the matrices $F_{k}^s$, $G_{k}^s$,  and $H_{k}^s$. This is done by taking the average of each uncertain position present in rows $i$ along the corresponding columns $l$ of matrices $F_{k}^s$, $G_{k}^s$, and $H_{k}^s$; see \cite{Inoue2016}. The obtained matrices are as follows:
\begin{equation}
\begin{array}{ccc}
N_{F_{k}}^s =10^{2}\begin{bmatrix} f_{1} ^s& \ldots & f_{n_{s}}^s\end{bmatrix},\\ N_{G_{k}}^s = 10^{2}\begin{bmatrix}g_{1} ^s& \ldots & g_{m_{s}}^s\end{bmatrix}, \\
N_{H_{k}}^s =\begin{bmatrix} 0 & \ldots & 0 \end{bmatrix}, N_{K_{k}}^s =\begin{bmatrix} 0 & \ldots &0\end{bmatrix},
\end{array}
\end{equation}
where $n_{s}$ is the number of variables in the state vector $\textbf{x}^s$; $m_{s}$ is the of variables in the noise vector $\textbf{w}^s$; $n_{a} = 6$; $m_{a} = 6$; $n_{p} = 9$; $m_{p} = 6$; $f_{l}^{s} = \frac{\sum_{i=1}^{n_{s}} \overline{F}_{k}^s(i,l)}{n_{s}}$, $\overline{F}_{k} ^s= F_{k} ^s- I_{n_{s}\times n_{s}}$, for $l = 1,2,..,n_{s}$; $g_{l}^{s} = \frac{\sum_{i=1}^{n_{s}} G_{k}^p(i,l)}{n_{s}}$, for $l = 1,\ldots~,m_{s}$.



\subsection{Plot of Attitude and Position Estimates} 


In this section, the graphs for the attitude and position estimates are presented. Figure \ref{fig:Euler_angles} shows the attitude estimates while Figure \ref{fig:position} shows the position estimates.

\subsubsection{Attitude Estimates}

\begin{figure*}[!ht]	
	\centering{
		\begin{minipage}{0.31\textwidth}
			\centering
			\includegraphics[width=1\textwidth]{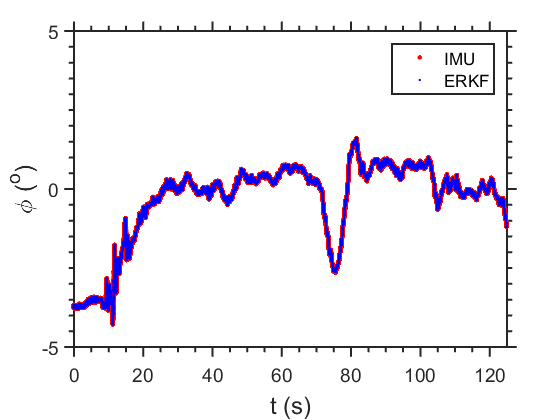}\label{fig:phi} 
		\end{minipage}
		\begin{minipage}{0.31\textwidth}
			\centering
			\includegraphics[width=1\textwidth]{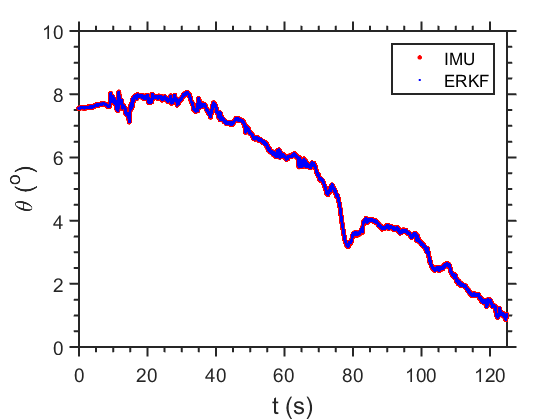}\label{fig:theta} 
		\end{minipage}
		\begin{minipage}{0.31\textwidth}
			\centering
			\includegraphics[width=1\textwidth]{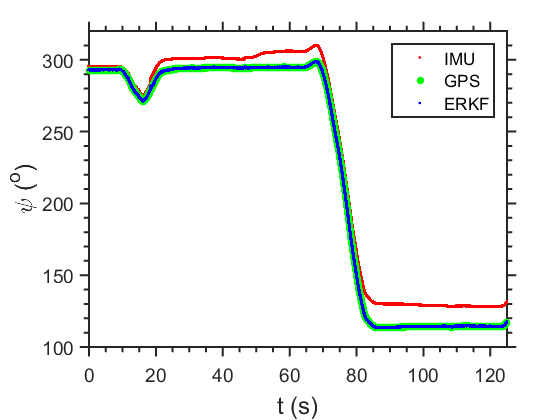}\label{fig:psi} 
		\end{minipage}
	}
	\caption{Attitude estimates of the ERKF: roll ($\phi$),  pitch ($\theta$) and  yaw ($\psi$).}
	\label{fig:Euler_angles}
\end{figure*}

The majority of variation in attitude is seen in the yaw angle as shown in Figure \ref{fig:Euler_angles}(c), while the total variation in roll, Figure \ref{fig:Euler_angles}(a), and pitch, Figure \ref{fig:Euler_angles}(b), are considerably lower. This is as expected for a ground vehicle. 

The yaw angle estimates provided by the filter, Figure \ref{fig:Euler_angles}(c), show greater certainty in GPS measurements by following them more closely than IMU measurements, which are observed to have errors due to the presence of uncertainties. It should be noted that due to the ERKF the yaw estimates are robust to the uncertainties present in the IMU measurements.


\subsubsection{Position Estimates}

\begin{figure}[ht]
	\centering
	\includegraphics[width=0.85\linewidth]{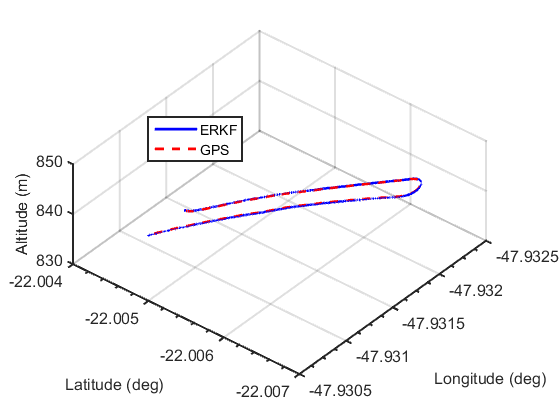} 
	\caption{3D Plot of Vehicle Position.} \label{route_view}
\end{figure}

\begin{figure}[ht]
	\centering
	\includegraphics[width=0.85\linewidth]{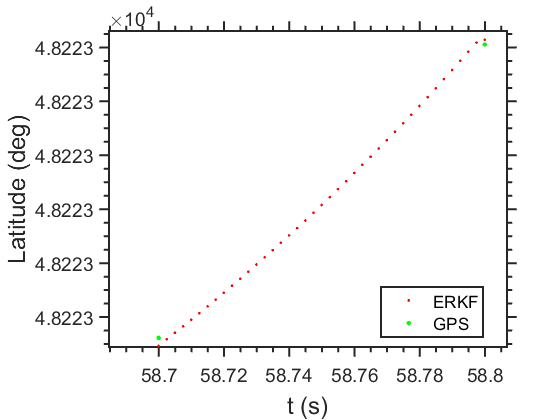} 
	\caption{Update Rates of GPS and Estimates.} \label{zoom}
\end{figure}

\begin{figure*}[!h]	
	\centering{
		\begin{minipage}{0.31\textwidth}
			\centering
			\includegraphics[width=1\textwidth]{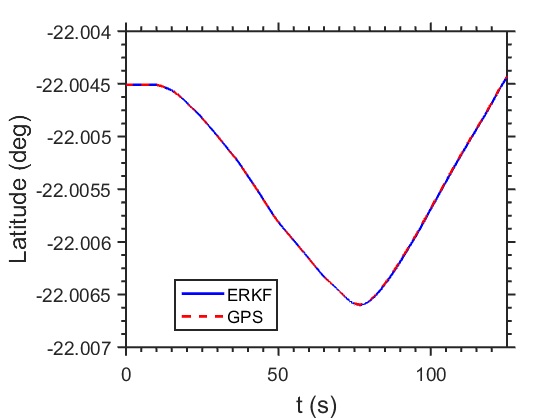}\label{fig:latitude} 
		\end{minipage}
		\begin{minipage}{0.31\textwidth}
			\centering
			\includegraphics[width=1\textwidth]{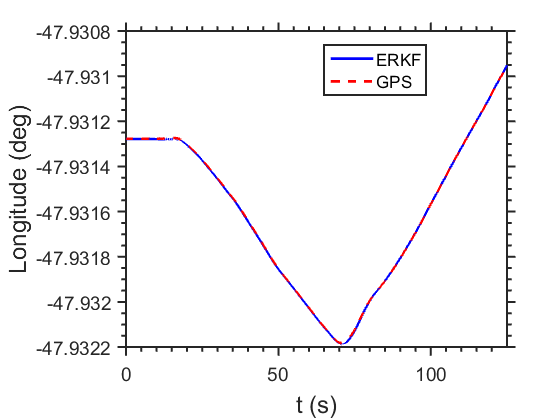}\label{fig:longitude} 
		\end{minipage}
		\begin{minipage}{0.31\textwidth}
			\centering
			\includegraphics[width=1\textwidth]{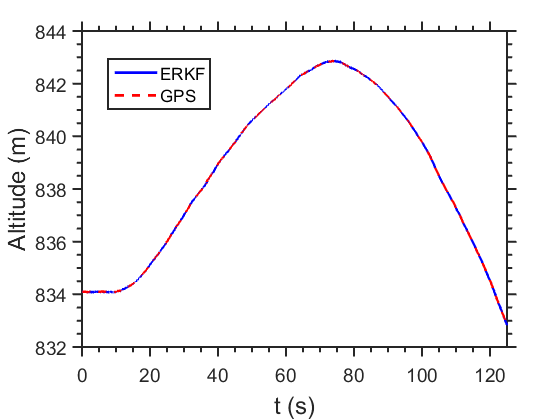}\label{fig:altitude} 
		\end{minipage}
	}
	\caption{Position estimates of the ERKF: latitude, longitude and  altitude.}
	\label{fig:position}
\end{figure*}

The position estimates are a good fit with GPS readings as shown in Figure \ref{fig:position} as well as in Figure \ref{route_view}, which is a 3 dimensional plot showing the route followed by the cargo transport vehicle. An important observation to be made here is that by implementing an inertial navigation system when GPS measurements are not available, the position estimates are updated at a frequency of 400 Hz, which is greater than the 10 Hz update rate of the GPS. This is depicted in Figure \ref{zoom}.

\subsection{Numerical analysis} \label{sec:numerical_analisys}

\begin{figure*}[!htb]
	\centering{
		\begin{minipage}{0.31\textwidth}
			\centering
			\includegraphics[width=1\textwidth]{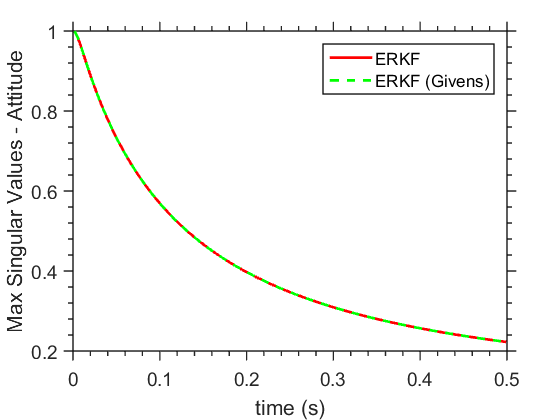}\label{fig:SVD_max_attitude}
		\end{minipage}
		\begin{minipage}{0.31\textwidth}
			\centering
			\includegraphics[width=1\textwidth]{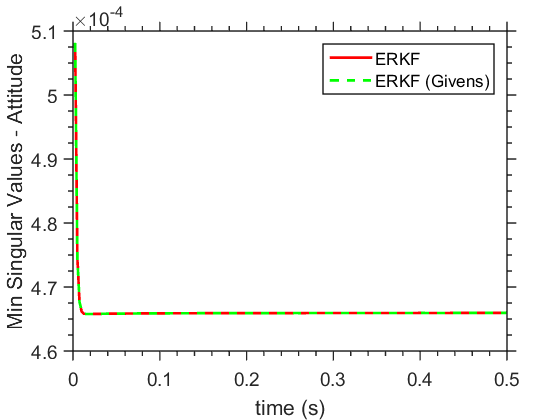}\label{fig:SVD_min_attitude}
		\end{minipage}
	}
	\caption{Maximum and minimum singular values of state covariance matrix $P$ of the attitude system model.}
	\label{fig:SVD_attiude_system}
\end{figure*}

\begin{figure*}[!htb]
	\centering{
		\begin{minipage}{0.31\textwidth}
			\includegraphics[width=1\textwidth]{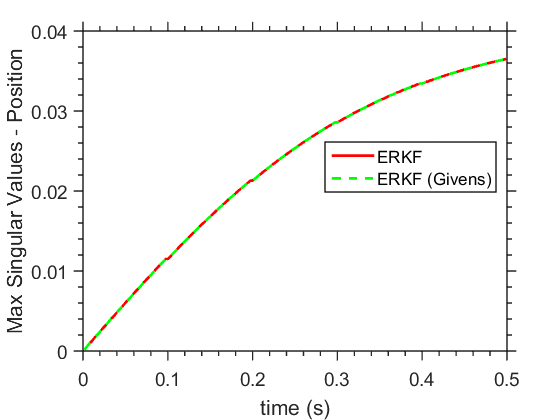}\label{fig:SVD_max_position}
		\end{minipage}
		\begin{minipage}{0.31\textwidth}
			\includegraphics[width=1\textwidth]{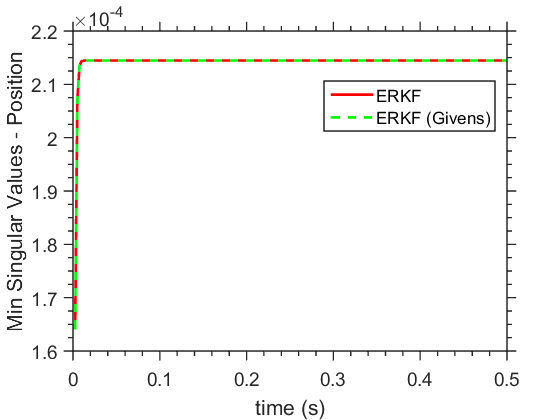}\label{fig:SVD_min_position}
		\end{minipage}
	}
	\caption{Maximum and minimum singular values of state covariance matrix $P$ of the position system model.}
	\label{fig:SVD_position_system}
\end{figure*}

The maximum and minimum singular values of the state covariance matrix of the attitude system are shown in Figure \ref{fig:SVD_attiude_system}. The ERKF has been implemented through the conventional matrix inversion approach as well as the proposed Givens rotation approach. The corresponding maximum and minimum singular values of the state covariance matrix of the position system are shown in Figure \ref{fig:SVD_position_system}.

These figures clearly show that the singular values of the covariance matrices obtained from the standard matrix inversion and Givens rotation implementations are very nearly the same. It was seen that absolute value of the difference between the singular values was smaller than $10^{-13}$ when 64 bit precision floating point arithmetic was used.


\section{Conclusions} \label{sec:conclu}

In this paper we have presented an attitude and heading reference system, based on IMU and GPS data, using the ERKF. The yaw estimates provided by the ERKF followed the accurate GPS measurements. The position estimates were obtained at a higher update frequency, due to the utilization of an inertial navigation system based on attitude estimates. And the ERKF ensured that the estimates are robust to uncertainties in both system models.

Additionally, we have presented and verified a method for computing the ERKF in real-time based on QR decomposition using Givens rotation. The increased computational efficiency of this method over the conventional matrix inversion approach has been discussed.

\section*{Acknowledgement}
This work was supported by grants \#2014/08432-0, \#2014/50851-0 and \#2015/18085-8, S\~ao Paulo Research Foundation (FAPESP) and by grants \#484095/2013-7 and 465755/2014-3, Brazilian National Council for Scientific and Technological Development (CNPq).


\end{document}